\documentclass[12pt]{article}
\usepackage{bm}
\usepackage{latexsym}
\usepackage{amsthm,amsfonts,color,colordvi,url}
\usepackage[]{graphicx}

\expandafter\ifx\csname package@font\endcsname\relax\else
 \expandafter\expandafter
 \expandafter\usepackage
 \expandafter\expandafter
 \expandafter{\csname package@font\endcsname}
\fi

\newcommand{\maj}{\text{Maj}}

\newcommand{\pbfrac}[2]{\mbox{$\mbox{}^{#1}\!/_{#2}$}}

\newtheorem{theo}{Theorem}
\newtheorem{theorepeat}{Theorem}
\theoremstyle{definition}
\newtheorem{defi}{Definition}
\newtheorem{lemma}{Lemma}
\newtheorem{corol}{Corollary}
\newtheorem{rem}{Remark}

\begin{document}

\title{A limit on nonlocality in any world in which communication complexity is not trivial}

\newcommand{\email}[1]{}
\newcommand{\affiliation}[1]{}
\newcommand{\pacs}[1]{}
\newcommand{\keywords}[1]{\vspace{2ex}\noindent\textbf{Keywords:} #1}
\newcommand{\text}[1]{{\mbox{{#1}}}}

\author{Gilles Brassard$^{\,1}$,
Harry Buhrman$^{\,2,\,3}$,
Noah Linden$^{\,4}$ \\
Andr\'e Allan M\'ethot$^{\,1}$,
Alain Tapp$^{\,1}$ and
Falk Unger$^{\,3}$
\\ \small 1. \sl D\'epartement IRO, Universit\'e de Montr\'eal,
Canada
\\[-1ex] \phantom{\small 1.} \texttt{\small\{brassard,\,methotan,\,tappa\}@iro.umontreal.ca}
\\ \small 2. \sl ILLC, Universiteit van Amsterdam,
The Netherlands
\\ \small 3. \sl Centrum voor Wiskunde en Informatica (CWI),
Amsterdam, The Netherlands
\\[-1ex] \phantom{\small 3.} \texttt{\small\{harry.buhrman,\,falk.unger\}@cwi.nl}
\\ \small 4. \sl Department of Mathematics, University of Bristol,
United Kingdom
\\[-1ex] \phantom{\small 4.} \texttt{\small n.linden@bristol.ac.uk}
}

\date{4 August 2005}

\maketitle


\begin{abstract}
Bell proved that quantum entanglement enables two space-like
separated parties to exhibit classically impossible correlations.
Even though these correlations are stronger than
anything classically achievable, they cannot be harnessed to make
instantaneous (faster than light) communication possible.
Yet, Popescu and Rohrlich have shown that even stronger correlations can be defined,
under which instantaneous communication remains impossible.
This raises the question: Why are the corre\-la\-tions achievable by quantum
mechanics not maximal among those that preserve causality?
We~give a partial answer to this question by showing that slightly
stronger correlations would result in a world in which communication
complexity becomes trivial.
\end{abstract}



\section{Introduction}
In the field of quantum information processing, entanglement can be
harnessed to accomplish amazing feats, such as quantum
teleportation~\cite{bbcjpw93}. The first proof that genuinely
nonclassical behaviour could be produced by quantum-mechanical
devices was given by John Bell in 1964~\cite{bell64}, when he proved
that quantum entanglement enables two space-like separated parties
to exhibit correlations that are stronger than anything allowed by
classical physics.  A~few years later, John Clauser, Michael Horne,
Abner Shimony and Richard Holt (CHSH), inspired by the work of Bell,
proposed another inequality~\cite{chsh69}, which was easier to
translate into a feasible experiment~\cite{clauser,aspect} to test
local hidden-variable theories. Their proposal fits nicely into the
more modern framework of nonlocal boxes, introduced by Sandu Popescu
and Daniel Rohrlich~\cite[Eq.~(7)]{pr94}.

A \emph{nonlocal box} (NLB) is an imaginary device that has an
input-output port at Alice's and another one at Bob's, even though
Alice and Bob can be space-like separated. Whenever Alice feeds a
bit $x$ into her input port, she gets a uniformly distributed random
output bit $a$, locally uncorrelated with anything else, including
her own input bit.  The same applies to~Bob, whose input and output
bits we call $y$ and $b$, respectively. The~``magic'' appears in the
form of a correlation between the pair of outputs and the pair of
inputs: the exclusive-or (sum modulo two, denoted~``$\oplus$'') of
the outputs is always equal to the logical \textsc{and} of the
inputs: \mbox{$a\oplus b = x \wedge y$}. Much like the correlations
that can be established by use of quantum entanglement, this device
is atemporal: Alice gets her output as soon as she feeds in her
input, regard\-less of if and when Bob feeds in \emph{his} input,
and vice versa. Also inspired by entanglement, this is a
\emph{one-shot} device: the~correlation appears only as a result of
the first pair of inputs fed in by Alice and Bob, respectively.
Of~course, they can have more than one NLB at their disposal, which
is then seen as a \emph{resource} of a different nature than
entanglement~\cite{bm05}.

A crucial property of NLBs is that they cannot be used by Alice and Bob
to signal instantaneously (faster than light) to one another. This is because the outputs
that can be observed are purely random from a local perspective. In~other words,
NLBs are nonlocal, yet they are \emph{causal}: they cannot make an effect
precede its cause in the context of special relativity.
We~are interested in the question of how well the correlation of NLBs can be
\emph{approxi\-mated} by devices that follow the laws of physics.

Even though originally presented in different terms,
it is easy to recast the CHSH inequality in the language of imperfect~NLBs.
The availability of prior shared entanglement allows Alice and Bob to
approximate NLBs with a success probability equal to
\[ \textstyle
\wp = \cos^2 \frac{\pi}{8} = \frac{2+\sqrt{2}}{4} \approx 0.854 \, .
\]
This can be used to test local hidden-variable theories because it
follows also from CHSH that no local realistic (classical) theory
can succeed with probability greater than~$\pbfrac34$ if Alice and
Bob are space-like separated. Later, Boris Tsirelson proved the
optimality of the CHSH inequality, which translates into saying that
quantum mechanics does not allow for a success prob\-ability greater
than $\wp$ at the game of simulating~NLBs~\cite{cirelson80}. See
also \cite{BuhrMass04} for an information-theoretic proof of the
same result.

The questions of interest in this paper are: (1)~\textsl{Considering
that perfect NLBs would not violate causality, why do the laws of
quantum mechanics only allow us to implement NLBs better than
anything classically possible, yet not perfectly?}; and
(2)~\textsl{Why do they provide us with an approximation of NLBs
that succeeds with probability $\wp$ rather than something better?}

Before we can pursue this line of thought further, we need to review
briefly the field of (\emph{quantum}) \emph{communication
complexity}~\cite{NisKus,CleveBuhrman,survey2,survey}. Assume Alice
and Bob wish to compute some Boolean function \mbox{$f(x,y)$} of
input $x$, known to Alice only, and input $y$, known to Bob only.
Their concern is to minimize the amount of (classical) communication
required between them for Alice to learn the answer. It~is clear
that this task cannot be accomplished without at least \emph{some}
communication (even if Alice and Bob share prior entanglement),
unless \mbox{$f(x,y)$} does not actually depend on~$y$, because
otherwise instantaneous signalling would be possible. Thus, we say
that the communication complexity of $f$ is \emph{trivial} if the
problem can be solved with a \emph{single bit} of communication.

It~is known that prior entanglement shared between Alice and Bob
helps sometimes but not always. Some functions can be computed with
exponentially less communication than would have been required in a
purely classical world~\cite{BCW}.  Yet other functions, such as the
\emph{inner product}, require as many bits to be communicated as the
length of Bob's input, whether or not prior entanglement is
available~\cite{cdnt97}: those are not trivial. Surprisingly,
Wim~van~Dam~\cite{vandam}, and independently Richard
Cleve~\cite{Cleve}, proved that the availability of perfect NLBs
makes the communication complexity of \emph{all} Boolean functions
trivial! This answers the first question above: If~we take as an
axiom that communication complexity should not be trivial in the
real world, it had to be impossible for quantum mechanics to provide
a perfect implementation of~NLBs. Indeed, most computer scientists
would consider a world in which communication complexity is trivial
to be as surprising as a modern physicist would find the violation
of causality.

In~order to answer the second question, we turn our attention to the \emph{probabilistic}
version of communication complexity, in which we do not require Alice to learn the
value of \mbox{$f(x,y)$} with certainty.  Instead, we shall be satisfied if she can obtain
an answer that is correct with a probability bounded away from~$\pbfrac12$. In~other words,
there must exist some real number \mbox{$p>\pbfrac12$} such that the probability that Alice
guesses the correct value of \mbox{$f(x,y)$} is at least~$p$ for \textit{all} pairs
\mbox{$(x,y)$} of inputs. Here, the probability is taken over possible probabilistic
behaviour at Alice's and Bob's, as well as over the value of random variables shared between
Alice and Bob. Note that there is no need for Boolean shared random variables if prior
entanglement or perfect NLBs is available since either one of those resources can be used by
Alice and Bob to obtain identical yet random coin tosses.

When we extend the notion of ``trivial'' communication complexity to
fit this probabilistic framework, the computation of the inner
product remains nontrivial according to quantum mechanics: Even if
Alice and Bob share prior entanglement, they need to transmit at
least \mbox{$\max(\frac12 (2p-1)^2,(2p-1)^4)n-\frac12$} classical
bits in order to succeed with probability at least \mbox{$p >
\pbfrac12$}, which is linear in the length $n$ of the inputs when
$p$ is a constant~\cite{cdnt97}. Our main theorem, stated below and
proven in Section~\ref{sc:main}, provides a partial answer to the
second question.

\begin{theo}\label{th:main}
In any world in which it is possible, without communication,
to~implement an approximation to the NLB that works correctly with
prob\-ability greater than \mbox{$\frac{3+\sqrt{6}}{6} \approx
90.8\%$}, \emph{every} Boolean function has trivial probabilistic
communication complexity.
\end{theo}

To prove this theorem, we introduce the notion of \emph{distributed computation}
and the notion of \emph{bias} for such computations.
Then, we show how to amplify the natural bias of \emph{any} Boolean function
by having Alice and Bob calculate it many times and taking the majority.
We~deter\-mine how imperfect a majority gate can be and still increase the bias.
Finally, we construct a majority gate with the use of NLBs, and
we determine to what extent we can allow \emph{them} to be faulty.


\section{Preliminary Definitions and Lemmas}

\begin{defi}
A~bit $c$ is \emph{distributed} if Alice has bit $a$ and
Bob bit $b$ such that $c = a \oplus b$.
\end{defi}

\begin{defi}
A Boolean function $f$ is \emph{distributively computed} by Alice and Bob
if, given inputs $x$ and $y$, respectively, they can produce a distributed bit
equal to \mbox{$f(x,y)$}.
Communication is not allowed during a distributed computation.
\end{defi}

\begin{defi}
A Boolean function is \emph{biased} if it can be distributively computed
with probability strictly greater than~$\pbfrac12$.
\end{defi}

\begin{lemma}\label{ntb}
Provided Alice and Bob are allowed to share random variables, \emph{all} Boolean functions are biased.
\end{lemma}

\begin{proof}
Let $f$ be an arbitrary Boolean function and let Alice and Bob share
a uniformly distributed random variable $z$ of the same size as
Bob's input~$y$. (It~is usual practice in communication
complexity~\cite{NisKus} to assume that each party knows the size of
both inputs.) Upon receiving her input $x$, Alice produces
\mbox{$a=f(x,z)$}. Bob's strategy is to test if \mbox{$y=z$}. If~so,
he produces \mbox{$b=0$}; if not, he produces a uniformly
distributed random bit~$b$. In~the lucky event that \mbox{$y=z$},
the bit distributed between Alice and Bob is \mbox{correct} since
\mbox{$a \oplus b = f(x,z) \oplus 0 = f(x,y)$}. This happens with
probability $2^{-n}$ if $n$ is the size of Bob's input. In~all other
cases (with overwhelming probability \mbox{$1-2^{-n}$}), the
distributed bit \mbox{$a \oplus b$} is uniformly random, hence it is
correct with probability~$\pbfrac12$. Summing up, the distributed
bit is correct with probability
\[
\Pr[a \oplus b = f(x,y)] = \frac{1}{2^n} +
\Bigl(1-\frac{1}{2^n}\Bigr)\frac12 = \frac12 + \frac{1}{2^{n+1}} \,
,
\]
which is indeed strictly greater than~$\pbfrac12$.
\end{proof}

\begin{defi}
A Boolean function has \emph{bounded bias} if it can be distributively computed
with probability bounded away from~$\pbfrac12$.
\end{defi}

\begin{rem}
The difference between bias and \emph{bounded} bias is that the
probability of being correct in the former case can come arbitrarily
close to $\pbfrac12$ as the size of the inputs increases. In~the
latter case, there must be some fixed \mbox{$p>\pbfrac12$} such that
the probability of being correct is at least $p$ no matter how large
the inputs are.
\end{rem}

\begin{lemma}\label{lem:bias}
Any Boolean function that has bounded bias has trivial probabilistic communication complexity.
\end{lemma}

\begin{proof}
Assume Boolean function $f$ has bounded bias.
For all inputs $x$ and $y$, Alice and Bob can produce bits $a$ and $b$, respectively,
without communication, such that \mbox{$a \oplus b=f(x,y)$} with probability at least
\mbox{$p > \pbfrac12$}.
If~Bob sends $b$ to Alice, she can compute \mbox{$a \oplus b$},
which is equal to \mbox{$f(x,y)$} with bounded error probability,
after a single bit has been communicated.
\end{proof}

\begin{defi}
The \emph{nonlocal majority} problem consists in computing the distributed majority of
three distributed bits. More precisely, let Alice have bits $x_1$, $x_2$, $x_3$ and Bob
have $y_1$, $y_2$,~$y_3$. The~purpose is for Alice and Bob to compute $a$ and~$b$,
respectively, such that
\[
a \oplus b = \maj(x_1 \oplus y_1, x_2 \oplus y_2, x_3 \oplus y_3) \, ,
\]
where $\maj(u,v,w)=\lfloor(u+v+w)/2\rfloor$ denotes the most
frequent bit among $u$, $v$ and~$w$. The computation of $a$ and $b$
must be achieved without any communication between Alice and Bob.
\end{defi}

John von Neumann proved a statement rather similar to Lemma~\ref{lem:jvn} below in 1956,
but in the context of ordinary circuits rather than distributed computation~\cite{vNeumann}.
We~sketch the proof nevertheless for the sake of completeness.

\begin{lemma}\label{lem:jvn}
For any $q$ such that \mbox{$\pbfrac56 < q \leq 1$}, if Alice and Bob
can compute nonlocal majority with probability at least $q$,
every Boolean function has bounded bias.
\end{lemma}

\begin{proof}
Let $f$ be an arbitrary Boolean function, fix Bob's input size, and
consider any \mbox{$p>\pbfrac12$} so that Alice and Bob can
distributively compute $f$ with probability at least~$p$. We~know
from Lemma~\ref{ntb} that such a $p$ exists (although it can depend
on the size of the inputs). Let Alice and Bob apply their
distributed computational process three times, with independent
random choices and shared random variables each time. This produces
three distributed bits such that each of them is correct with
probability at least~$p$. Let now Alice and Bob compute the nonlocal
majority of these three outcomes with probability at least $q$ that
the nonlocal majority be computed correctly. Because the overall
result will be correct either if most of the distributed outcomes
were correct and the distributed majority calculation was performed
correctly, or if most of the distributed outcomes were wrong and the
distributed majority calculation was performed incorrectly, the
probability that the distributed majority as computed yields the
correct value of $f$ is
\[
h(p) = q(p^3 + 3p^2(1-p)) + (1-q)(3p(1-p)^2+(1-p)^3) \, .
\]
Define
\[
\delta = q - \pbfrac56 > 0 \text{~~and~~}
s = \frac12 + \frac{3\sqrt{\delta}}{2\sqrt{1+3 \delta}} > \frac12 \, .
\]
It can be shown that \mbox{$p < h(p) < s$} provided \mbox{$\pbfrac12
< p < s$}. Because of this and the fact that $h(p)$ is continuous
over the entire range \mbox{$\pbfrac12 < p < s$}, iteration of the
above process can boost the probability of distributively computing
the correct answer arbitrarily close to~$s$. This proves that $f$
has bounded bias because, given any fixed value of
\mbox{$q>\pbfrac56$}, we can choose an arbitrary constant \mbox{$t <
s$} such that \mbox{$t > \pbfrac12$} and distributively compute $f$
with probability at least $t$ of being correct, independently of the
size of the inputs.
\end{proof}

\begin{defi}
The \emph{nonlocal equality} problem consists in distributively
deciding if three distributed bits are equal. More precisely, let
Alice have bits $x_1$, $x_2$, $x_3$ and Bob have $y_1$,
$y_2$,~$y_3$. The~purpose is for Alice and Bob to compute $a$
and~$b$, respectively, such that
\[
a \oplus b =
\left\{
\begin{array}{ll}
1 & \text{if } x_1 \oplus y_1 = x_2 \oplus y_2 = x_3 \oplus y_3 \\[1ex]
0 & \text{otherwise} \, .
\end{array}
\right.
\]
The computation of $a$ and $b$ must be achieved without any communication
between Alice and Bob.
\end{defi}

\begin{lemma}\label{lem:NLE}
Nonlocal equality can be computed using only two (perfect) nonlocal boxes.
\end{lemma}

\begin{proof} The goal is to obtain $a$ and $b$ such that:
\begin{eqnarray}
\label{eq1} a \oplus b &\!\!=\!\!& (x_1 \oplus y_1 = x_2 \oplus y_2)
\wedge (x_2 \oplus y_2 = x_3 \oplus y_3) .
\end{eqnarray}
First, Alice and Bob compute locally
\[ x'= \overline{x_1} \oplus x_2 ,~~ y'=y_1 \oplus y_2 ,~~ x''= \overline{x_2} \oplus x_3 \text{~~and~~}
   y''=y_2 \oplus y_3 \, . \]
Then (\ref{eq1}) becomes equivalent to \mbox{$(x'\oplus y') \wedge
(x'' \oplus y'')= a \oplus b$}. So, it is sufficient to show how
Alice and Bob can compute the AND of the distributed bits
\mbox{$x'\oplus y'$} and \mbox{$x'' \oplus y''$}.

By the distributivity law of the AND over the exclusive-or, we have
\[ (x'\oplus y') \wedge (x'' \oplus y'') =
   (x' \wedge x'') \oplus (x'\wedge y'') \oplus (x''\wedge y') \oplus (y'\wedge y'') \, . \]
Using two nonlocal boxes, Alice and Bob can compute distributed bits \mbox{$a'\oplus b'$} and
\mbox{$a''\oplus b''$} with \mbox{$a'\oplus b'=x'\wedge y''$} and
\mbox{$a''\oplus b'' = x''\wedge y'$}.
Setting \mbox{$a= (x'\wedge x'') \oplus a' \oplus a''$} and
\mbox{$b= (y'\wedge y'') \oplus b' \oplus b''$} yields (\ref{eq1}), as desired.
\end{proof}

\begin{lemma}\label{lem:NLM}
Nonlocal majority can be computed using only two (perfect) nonlocal boxes.
\end{lemma}

\begin{proof}
Let $x_1$, $x_2$, $x_3$ be Alice's input and $y_1$, $y_2$, $y_3$ be Bob's.
For \mbox{$i \in \{1,2,3\}$}, let \mbox{$z_i=x_i \oplus y_i$} be the
$i^\text{\scriptsize th}$ distributed input bit.
By virtue of Lemma~\ref{lem:NLE}, Alice and Bob use their two NLBs
to compute the nonlocal equality of their inputs, yielding
$a$ and $b$ so that \mbox{$a \oplus b=1$}
if and only if $z_1$, $z_2$ and $z_3$ are equal.
Finally, Alice produces \mbox{$a' = \overline{a} \oplus x_1 \oplus x_2 \oplus x_3$}
and Bob produces \mbox{$b' = b \oplus y_1 \oplus y_2 \oplus y_3$}.
Let
\[ z = a' \oplus b' = (\overline{a} \oplus b) \oplus (z_1 \oplus z_2 \oplus z_3) \]
be the distributed bit computed by this protocol. Four cases need to
be considered, depending on the number $\ell$ of 1s among the
$z_i$'s:
\begin{enumerate}
    \item if $\ell=0$, then $a \oplus b = 1$ and $z_1 \oplus z_2 \oplus z_3 = 0$;
    \item if $\ell=1$, then $a \oplus b = 0$ and $z_1 \oplus z_2 \oplus z_3 = 1$;
    \item if $\ell=2$, then $a \oplus b = 0$ and $z_1 \oplus z_2 \oplus z_3 = 0$;
    \item if $\ell=3$, then $a \oplus b = 1$ and $z_1 \oplus z_2 \oplus z_3 = 1$.
\end{enumerate}
We see that \mbox{$z=0$} in the first two cases and \mbox{$z=1$} in
the last two, so that \mbox{$z=\maj(z_1,z_2,z_3)$} in all cases.
\end{proof}

We are now ready to prove our main theorem.

\section{Proof of the Main Theorem}\label{sc:main}

Before proving it, let us repeat the statement of the main theorem.

\begin{theorepeat}
In any world in which it is possible, without communication,
to~implement an approximation to the NLB that works correctly with
prob\-ability greater than \mbox{$\frac{3+\sqrt{6}}{6} \approx
90.8\%$}, \emph{every} Boolean function has trivial probabilistic
communication complexity.
\end{theorepeat}

\begin{proof}
Assume NLBs can be approximated with some probability $p$ of yielding the correct result.
Using them, we can compute nonlocal majority with probability \mbox{$q=p^2+(1-p)^2$}
since the protocol given in the proof of Lemma~\ref{lem:NLM} succeeds
precisely if none or both of the NLBs behave incorrectly.
The~result follows from Lemmas~\ref{lem:bias} and~\ref{lem:jvn}
because \mbox{$q>\pbfrac56$} whenever
\smash{\mbox{$p>\frac{3+\sqrt{6}}{6} \approx 0.908$}}.
A~more precise calculation, based on the proof of Lemma~\ref{lem:jvn},
shows that the two-party computation of \emph{any} Boolean function
can be achieved using a single bit of communication, with a probability
of correct answer arbitrarily close to
\[
\frac12 + \frac{\sqrt{3 p^2 - 3 p + \pbfrac14}}{2p-1}  \]
provided \smash{\mbox{$p>\frac{3+\sqrt{6}}{6}$}},
making it trivial by definition.
\end{proof}

\begin{corol}
In any world in which probabilistic communication complexity is nontrivial,
nonlocal boxes cannot be implemented without communication,
even if we are satisfied in obtaining the correct behaviour with probability
\mbox{$\frac{3+\sqrt{6}}{6} \approx 90.8\%$}.
\end{corol}

\begin{rem}
Neither nonlocal majority nor nonlocal equality can be solved
exactly with a single nonlocal box.
Otherwise, entanglement could approximate that NLB well enough
to solve the nonlocal majority problem with
probability \mbox{$\wp \approx 0.854 >\pbfrac56$} of being correct~\cite{chsh69}.
It~would follow from Lemmas~\ref{lem:bias} and~\ref{lem:jvn}
that all Boolean functions have trivial probabilistic communication complexity
according to quantum mechanics. But we know this not to be the case, in particular
for the inner product~\cite{cdnt97}.
\end{rem}

\begin{rem}
Our results also give bounds on the maximum admissible \mbox{error}
for elementary gates in fault-tolerant circuits: In the proof of
Lemma~\ref{lem:NLE}, we show how to simulate distributed AND-gates.
Using NLBs with the (quantum-mechanically achievable) correctness
probability $\wp$, such a distrib\-uted AND is correct with
probability $(1-\wp)^2 + \wp^2=\pbfrac34$. Further\-more, a
distributed NOT-gate on a distributed bit can be implemented
perfectly if just one party (say Alice) negates her bit. Like any
other Boolean function, the inner product
\mbox{\textit{IP}$(x,y)=\bigoplus_i (x_i \wedge y_i)$} can be
computed using only AND and NOT gates. But the entanglement-assisted
communication complexity of \textit{IP} is in $\Omega(n)$. Thus, if
we allow Alice and Bob to communicate only a constant number of bits
but allow them to use an arbitrary number of NLBs (with correctness
$\wp$), they still cannot compute the inner product function for
arbitrary inputs.

It follows that no family of circuits, consisting solely of
(perfect) NOT-gates and AND-gates that independently fail with
probability at most $\pbfrac14$, can compute the inner product
function for arbitrary input sizes. A~stronger result along these
lines had already been proven by William Evans and Nicholas
Pippenger~\cite{EvPip}, but their purely classical proof is
significantly more complicated.
\end{rem}


\section{Conclusions}

In conclusion, we have shown that in any world in which
communication complexity is nontrivial, there is a bound on how much
Nature can be nonlocal. For~this purpose, we developed a protocol to
distributively compute with bounded bias any Boolean function,
provided we can approximate nonlocal boxes with probability greater
than \smash{$\frac{3+\sqrt{6}}{6} \approx 0.908$}. This bound, which
is an improvement over previous knowledge that nonlocal boxes could
not be implemented exactly~\cite{vandam,Cleve}, approaches the
actual bound \mbox{$\wp \approx 0.854$} \mbox{imposed} by quantum
mechanics. The~obvious open question is to close the gap between
these probabilities. A~proof that nontrivial communication
complexity forbids nonlocal boxes to be approximated with
probability greater than~$\wp$ would be very interesting, as it
would make Tsirelson's bound~\cite{cirelson80} inevitable.
Conversely, if we could show how to use quantum entanglement to
approximate nonlocal equality with probability $\pbfrac56$ of
success, this would imply that our line of reasoning cannot be
improved.


\section*{Acknowledgements}
The authors are grateful to Richard Cleve for stimulating
discussions. H.B.,~N.L. and F.U. thank the Newton Institute where
part of this work was done. G.\,B.~is supported by the Natural
Sciences and Engineering Research Council of Canada ({\sc Nserc}),
the Canadian Institute for Advanced Research ({\sc Ciar}) and the
Canada Research Chair Programme. H.B., N.L. and F.U. are supported
by the EU under project RESQ (IST-2001-37559). H.B. and F.U. are
supported also by the NWO vici project 2004-2009. A.T.~is supported
by {\sc Nserc}, {\sc Ciar}, the Mathematics of Information
Technology and Complex Systems Network ({\sc Mitacs}) and the Fonds
Qu\'ebecois de Recherche sur la Nature et les Technologies ({\sc
Fqrnt}).


\end{document}